# Compacting Transactional Data in Hybrid OLTP&OLAP Databases

Florian Funke
TU München
Munich, Germany
florian.funke@in.tum.de

Alfons Kemper
TU München
Munich, Germany
alfons.kemper@in.tum.de

Thomas Neumann
TU München
Munich, Germany
thomas.neumann@in.tum.de

## ABSTRACT

Growing main memory sizes have facilitated database management systems that keep the entire database in main memory. The drastic performance improvements that came along with these in-memory systems have made it possible to re-unite the two areas of online transaction processing (OLTP) and online analytical processing (OLAP): An emerging class of hybrid OLTP and OLAP database systems allows to process analytical queries directly on the transactional data. By offering arbitrarily current snapshots of the transactional data for OLAP, these systems enable real-time business intelligence.

Despite memory sizes of several Terabytes in a single commodity server, RAM is still a precious resource: Since free memory can be used for intermediate results in query processing, the amount of memory determines query performance to a large extent. Consequently, we propose the *compaction* of memory-resident databases. Compaction consists of two tasks: First, separating the mutable working set from the immutable "frozen" data. Second, compressing the immutable data and optimizing it for efficient, memory-consumption-friendly snapshotting. Our approach reorganizes and compresses transactional data online and yet hardly affects the mission-critical OLTP throughput. This is achieved by unburdening the OLTP threads from all additional processing and performing these tasks asynchronously.

## 1. INTRODUCTION

Modern in-memory database systems with high-performance transaction processing capabilities face a dilemma: On the one hand, memory is a scarce resource and these systems would therefore benefit from compressing their data. On the other hand, their fast and lean transaction models penalize additional processing severely which often prevents them from compressing data in favor of transaction throughput. A good example is the lock-free transaction processing model pioneered by H-Store/VoltDB [18, 30] that executes transactions serially on private partitions without any overhead from buffer management or locking. This model allows for record-breaking transaction throughput, but necessitates that all transactions execute quickly to prevent congestion in the serial execution pipeline.

As a result of this dilemma, OLTP engines often refrain from compressing their data and thus waste memory space. The lack of a compact data representation becomes even more impeding, when the database system is capable of running OLAP-style queries on the transactional data, like the HyPer system [19] or SAP HANA [11]. In this scenario, compression does not only reduce memory consumption, but also promises faster query execution [32, 1, 4, 14]. To facilitate efficient query processing directly on the transactional data, these hybrid OLTP & OLAP systems create snapshots of the transactional data, that should not be ignored in the context of space efficiency. Therefore we introduce the notion of *compaction*, a concept that embraces two mechanisms that serve a similar purpose:

- Compression of the data set to save storage space and speed-up query execution.
- Reorganization of the data set for efficient and memory-consumption friendly snapshotting.

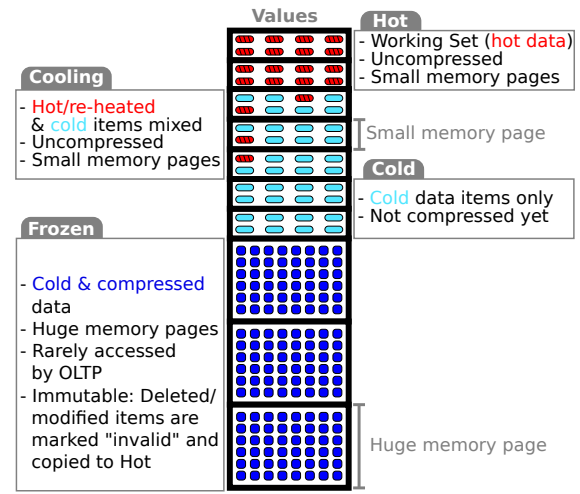

**Figure 1: Hot/cold clustering for compaction.**
🟥 = hot (volatile) data item, 🟦 = cold data item,
🔵 = cold & compressed data item.

We demonstrate that even though it is more difficult to compress transactional data due to its volatile nature, it is feasible to do it efficiently. Our approach is based on the observation that while OLTP workloads frequently modify the dataset, they often follow the working set assumption [10]: Only a small subset of the data is accessed and an even smaller subset of this working set is being modified (cf. Figure 1). In business applications, this working set is mostly comprised of tuples that were added to the database in the recent past, as it can be observed in the TPC-C workload [29].

Our system uses a lightweight, hardware-assisted monitoring component to observe accesses to the dataset and identify opportunities to reorganize data such that it is clustered into hot and cold parts. After clustering the data, the database system compresses cold chunks to reduce memory consumption and streamline query processing. Cold chunks are stored on huge virtual memory pages and protected from any modifications to allow for compact and fast OLAP snapshots. These physical reorganizations are performed at runtime with virtually no overhead for transaction processing as all time consuming tasks execute asynchronously to transaction processing.

The remainder of this paper is organized as follows: In Section 2, we give an overview of related work on compression for transaction processing systems and analytical systems. Section 3 presents our transaction model and physical data representation, how we cluster tuples into hot and cold parts and which compression schemes we propose to use. Furthermore, we describe the impact of our approach on query processing and present a space-efficient secondary index implementation. In Section 4, we describe techniques to observe data access patterns of transactions, including hardware-assisted, low-overhead mechanisms. Section 5 contains the experimental evaluation of our approach substantiating the claim that compaction hardly affects transaction processing while it can speed up query execution times. Section 6 concludes our findings.

## 2. RELATED WORK

Compression techniques for database systems is a topic extensively studied, primarily in the context of analytical systems [14, 32, 1, 16, 33, 28]. However, the proposed techniques are not directly applicable to OLTP systems which are optimized for high-frequency write accesses. We neither propose new compression algorithms, nor focus on the integration of compression with query processing in this work. Rather, we describe how existing compression techniques can be adapted for and integrated in transactional systems. The challenge in doing so is that compression must be performed in a non-disturbing manner with regard to the mission-critical transaction processing. In addition to OLTP, next-generation database systems like HyPer [19] and SAP HANA [11] offer OLAP capabilities operating directly on the transactional data. Thus, OLAP-style data accesses patterns must be taken into account when designing a compression features.

For efficient update handling in compressed OLAP databases, Héman et al. proposed Positional Delta Trees [15]. They allow for updates in ordered, compressed relations and yet maintain good scan performance. Binnig et al. propose ordered-dictionary compression that can be bulk-updated efficiently [3]. Both techniques are not designed for OLTP-style updates, but rather for updates in data warehouses. We will show how the benefits of order-preserving dictionary compression, which is infeasible in hybrid OLTP&OLAP systems, for query processing can be substituted efficiently in frequently changing datasets.

Oracle 11g [27] has an OLTP Table Compression feature. Newly inserted data is left uncompressed at first. When insertions reach a threshold, the uncompressed tuples are being compressed. Algorithm details or performance numbers are not published, but the focus appears to be on disc-based systems with traditional transaction processing models, not high-performance in-memory systems. Also, the feature seems to be applicable only in pure OLTP workloads without analytical queries.

Approaches that maintain two separate data stores, an uncompressed "delta" store for freshly inserted data and a compressed "main"-store for older data, require costly merge phases that periodically insert new data into the main store in a bulk operation [22]. This involves the exclusive locking of tables and also slows down transaction processing. Krüger et al. [21] state that their approach deliberately compromises OLTP performance to facilitate compression.

## 3. DESIGN

### 3.1 The HyPer System

We integrated our compaction approach into the HyPer [19] system, an in-memory, high-performance hybrid OLTP and OLAP DBMS. We briefly present the system here, but the approach is generally applicable to OLTP&OLAP in-memory database systems.

HyPer belongs to the emerging class of database systems that have – in addition to an OLTP engine – capabilities to run OLAP queries directly on the transactional data and thus enable real-time business intelligence. HyPer allows to run queries in parallel to transactions with extremely low overhead. This is achieved by executing the queries in a separate process that was `fork`ed from the OLTP process (cf. Figure 2) and constitutes a transaction-consistent snapshot of the OLTP data. The snapshot is kept consistent by the operating system with the assistance of the memory management unit (MMU). Before the OLTP process modifies a page, the original page is replicated for the OLAP process to maintain the memory state from the time of snapshot creation. This mechanism is a cornerstone of HyPer's performance, allowing it to compete with the fastest dedicated OLTP systems and the fasted dedicated analytical system – even when both workloads are executed in parallel on the same data in HyPer. Mühe et al. [25] found this hardware-based snapshotting approach superior to software-based solutions.

HyPer's transaction model is similar to the model pioneered by H-Store [18] and VoltDB [30]: The database is split into $p$ partitions, such that most transactions only need to access one of these partitions. Thus one OLTP-thread can be assigned to each partition and can operate within the partition without having to acquire any locks or latches. Only for the rare case of partition-crossing transactions, the $p$ threads must be synchronized. Moreover, transactions in HyPer are written as stored procedures and are compiled to native code by leveraging the LLVM compiler back-end, as described in [26]. This model allows for very fast transaction processing in the order of 100 000s of transactions per second [19].

HyPer's query engine is built upon a novel query compilation technique described in [26]. It avoids the performance

1425

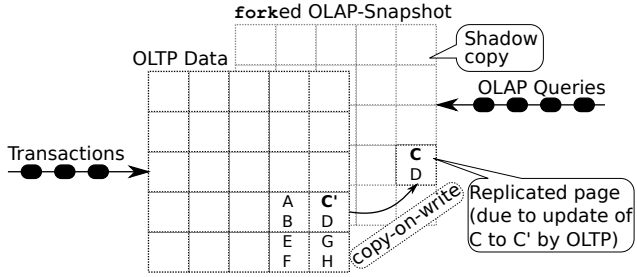

**Figure 2: HyPer's VM snapshot mechanism.** Only those memory pages get replicated that are modified during the lifetime of a snapshot.

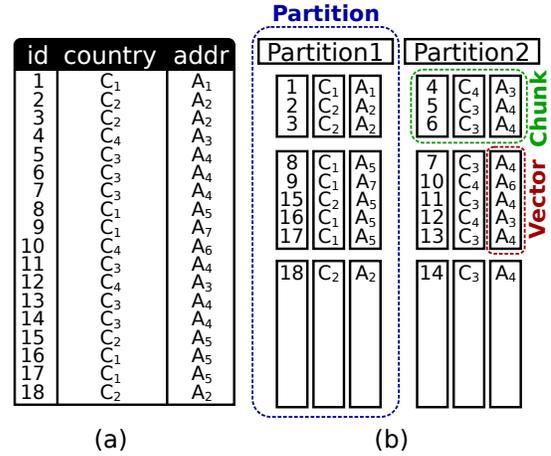

**Figure 3:** (a) Example relation. (b) Physical representation of example relation (without compression).

overhead of classic, iterator-style processing techniques that suffer from a lack of code locality and frequent instruction mispredictions by translating queries into LLVM code. This portable assembler code can be executed directly using LLVM's optimizing just-in-time compiler. It produces compact and efficient machine code that makes column scans completely data driven. Together with its sophisticated query optimizer, this enables HyPer to achieve sub-second query response times on typical business intelligence queries (top customers, top products, etc.) that can compete with the fastest dedicated OLAP systems.

### 3.2 Data Representation

We added a storage back-end to HyPer that combines horizontal partitioning and columnar storage: A relation is represented as a hierarchy of partitions, chunks and vectors (see Figure 3). Partitions split relations into $p$ disjoint subsets and are the basis of the transaction model described above. Within one partition, tuples are stored using a decomposed storage model [9]. Unlike designs where each attribute is stored in one continuous block of memory, we store a column in multiple blocks ("vectors"), as proposed in MonetDB/X100 [4]. In contrast to X100, our main rationale for doing so is that each vector that stores a certain attribute can represent it differently, e.g. compressed lightly, heavily or uncompressed. In addition, they can reside on different types of memory pages, i.e. regular or huge pages as discussed in Section 3.3. Each chunk constitutes a horizontal partition of the relation, i.e. it holds one vector for each of the relation's attributes and thus stores a subset of the partition's tuples, as depicted in Figure 3.

### 3.3 Hot/Cold Clustering

Hot/cold clustering aims at partitioning the data into frequently accessed data items and those that are accessed rarely (or not at all). This allows for physical optimizations depending on the access characteristics of data.

We measure the "temperature" of data on virtual memory page granularity. Since we store attributes column-wise, this allows us to maintain a separate temperature value for each attribute of a chunk, i.e. for each vector. Both read and write accesses to the vectors are monitored by the *Access Observer* component using a lightweight, hardware-assisted approach described in Section 4.1. It distinguishes four states a vector can have:

**Hot** Entries in a hot vector are frequently read/updated or tuples are deleted and/or inserted into this chunk.

**Cooling** Most entries remain untouched, very few are being accessed or tuples are being deleted in this chunk. If the Access Observer determines that the same pages were accessed between subsequent scans and they make up only a small fraction of the chunk's total number of pages, it marks the chunk cooling. Accessing an attribute in a cooling chunk triggers the relocation of the tuple to a hot chunk.

**Cold** Entries in a cold vector are not accessed, i.e. the Access Observer has repeatedly found no reads or writes in this vector.

**Frozen** Entries are neither read nor written and have been compressed physically and optimized for OLAP as described below.

"Access" refers only to reads and writes performed by OLTP threads – OLAP queries potentially executing in parallel do not affect the temperature of a vector, as described in Section 4.1.

Cold chunks of the data can be "frozen", i.e. converted into a compact, OLAP-friendly representation as they are likely to be almost exclusively accessed by analytical queries in the future. They are compressed, stored on huge virtual memory pages and made immutable. The use of huge pages (2MB per page on x86) for frozen data has multiple advantages over the use of regular pages (4kB on x86):

1. Scanning huge pages is faster than scanning regular pages. Manegold et al. [24] analyze the impact of translation lookaside buffer (TLB) misses on query performance and conclude with regard to the page size that the use of huge pages reduces the probability of TLB misses.

2. The TLB of the processor's memory management unit has separate sections for huge and normal pages on most platforms as depicted in Figure 4. Since the bulk of the database is frozen and resides on huge pages, table-scans in OLAP queries mostly access huge pages (and thus the TLB for huge pages). Transactions, on the other hand, operate almost entirely on hot data that is stored on normal pages. Thus, the two separate workloads utilize separate hardware resources and therefore do not compete for the TLB and do not cause "thrashing" of TLB.

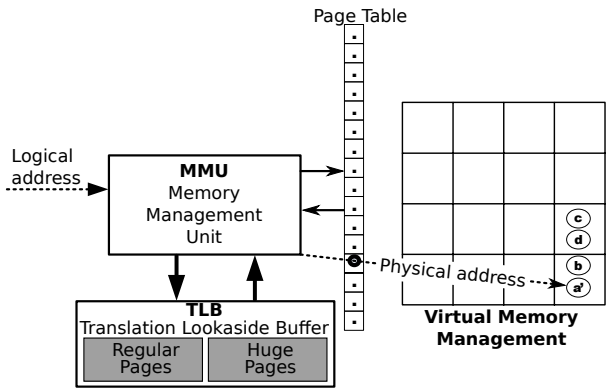

**Figure 4: Logical to physical address translation by MMU using its TLB.**

3. Huge pages speed up snapshotting. Snapshots facilitate the efficient execution of OLAP queries, even when transactions are being processed concurrently. Lorie's shadow paging [23] and its modern, hardware-assisted reincarnation, HyPer's `fork`-based snapshot mechanism, both benefit from the smaller page table size that results from the use of huge pages: The most time-consuming task the operating system has to perform when executing a `fork` is copying the process's page table. The use of huge pages can shrink the page table up to the factor 500 (on x86) and therefore facilitate considerably faster `fork`-times. Faster snapshotting in turn allows for more frequent snapshots. This does not only imply that queries see fresher data, but also that queries which require a new snapshot (e.g. to guarantee read-your-own-writes semantics) have a shorter delay.

Hot/cold clustering minimizes the memory overhead of snapshots that contain huge pages: Huge pages cannot be used for the entire database, but only for the immutable part, since all changes made by OLTP threads trigger the replication of that memory page when a snapshot is active. Because copying huge pages is significantly more expensive than copying regular pages, huge pages are only suitable for cold, read-only data in order to keep snapshots compact. This holds for HyPer's hardware-assisted snapshot mechanism as well as for the original software-based approach by Lorie. Even other software-based approaches to create consistent snapshots (see Mühe et al. [25] for a comparison), e.g. twin blocks [5], benefit from the hot/cold clustering as changes are concentrated in the hot part of the database.

Inserts are only made into chunks where all vectors are hot. If no hot chunk with available capacity exists, a new one is created. Updates and selects into hot chunks are simply executed in-place, while updates and selects in cooling chunks trigger the relocation of the tuple into a hot chunk to purge the cooling chunk from hot tuples. Deletes are carried out in-place in both hot and cooling chunks.

Updates and deletes are not expected in cold and frozen chunks. If they do occur, they are not executed in-place, but lead to the invalidation of the tuple and in case of an update also to its relocation to a hot chunk and an update in the index(es) that are used for point-accesses in OLTP transactions. This is depicted in Figure 5. We do so for two reasons: First, updating and deleting in-place can necessitate costly reorganizations. For example in run-length encoded vectors, splitting a run may requires up to two additional entries and thus force the vector to grow. Second, by refraining from updating compressed attributes in place, we keep snapshots compact, as huge pages are never written to and are thus never replicated. An invalidation status data structure is maintained to prevent table scans from passing the invalidated tuple to the parent operator. The invalidation status is managed following the idea of Positional Delta Trees [15]: The data structure records ranges of tuple IDs (TIDs) that are invalid and thus can be skipped when scanning a partition. We chose to record ranges of TIDs and not individual TIDs, because we expect that if updates/deletes happen to affect frozen chunks, they often occur in the following patterns:

- Very few updates/deletes affect a frozen chunk. In this case, the overhead of storing two values (range begin and range end) instead of a single value is very small.

- A large portion of the frozen data is being invalidated due to a change in the workload or administrative tasks. In this case, the data structure holds very few entries that specify to skip a very large portion of the data. In this case, storing the individual TIDs of invalidated tuples would cause overhead for scans and for memory consumption.

In addition to the aforementioned benefits of hot/cold clustering, separating the mutable from the immutable data items is advantageous for other components of the DBMS as well. As a frozen chunk is never modified in place, the recovery component can skip over all frozen chunks that have been persisted already, when it periodically writes a snapshot to disk (see [19] for details on HyPer's recovery component). Thus, for these chunks only the invalidation status has to be included when writing further snapshots to disk. This reduces the required disk IO of the recovery component significantly.

While transfering cold data to disk or SSD is possible in principle, discussing the implications is beyond the scope of this paper.

### 3.4 Compression

This paper focuses on the integration of existing compression schemes to high-performance OLTP systems with query capabilities. We do not propose new compression algorithms, but use well-known algorithms and apply them adaptively to parts of the data depending on the access patterns that are dynamically detected at runtime. It is our main goal to impact transaction processing as little as possible. Thus we release the transaction processing threads of all compression tasks. The small part of data that is frequently accessed – and freshly inserted data in particular – is left uncompressed and thus efficiently accessible for transactions. For cold chunks, we propose to use dictionary compression and run-length encoding, which was found to be beneficial for column stores [32, 1].

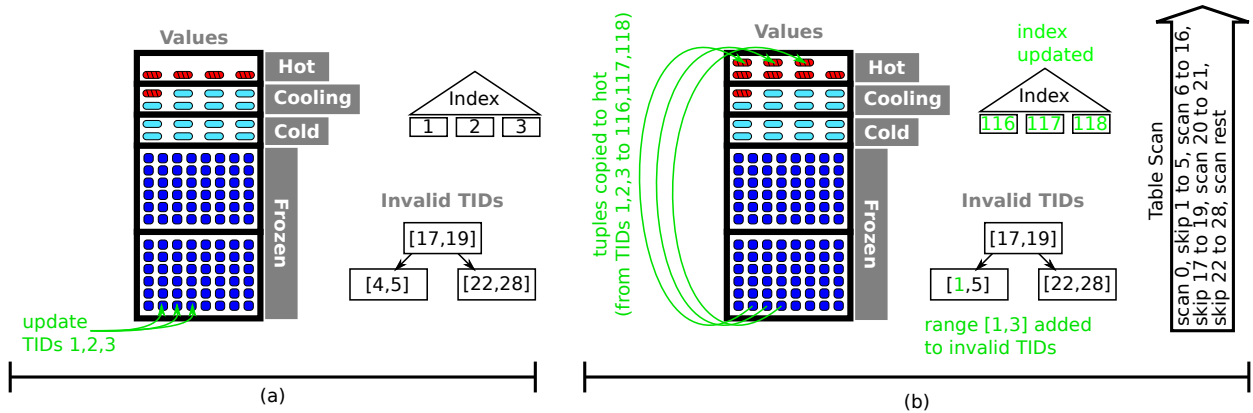

Figure 5: (a) Before an update of TIDs 1, 2 and 3. (b) After the update. Frozen chunks (residing on huge pages) are not modified, but invalidated out-of-place. Indexes (used by OLTP) are updated. Table scans "skip" over invalid ranges.

We propose to compress tuples once they stop being in the working set as opposed to compressing them when they are inserted for two reasons. First, compressing tuples on insert is more costly. When inserting into TPC-C's ORDERLINE relation, we measured a decline of the insert-rate to 50% when dictionary-compressing the single character attribute immediately (see Section 5 for details). Second, compressing only cold data allows to use a single dictionary for all partitions of a relation, without causing lock contention for the OLTP threads. A single dictionary has a much smaller memory footprint than $p$ dictionaries, especially since all $p$ dictionaries would presumably contain the same values and hence the memory consumption would be $p$ times higher than necessary is not unlikely. It thus yields much better compression rates and faster query execution time, since each table scan only involves a single dictionary scan.

Where beneficial, run-length encoding (RLE) is used on top of the dictionary compression. This further shrinks the size of the dictionary-key columns and improves scan performance. Examples where the use of RLE is advantageous include attributes that contain many null values or date fields that are set to the current date when inserting a tuple.

While frozen chunks are designed for efficient scan access and not point accesses, point accesses made by transactions must still be possible with acceptable performance – even if they are expected to occur rarely in frozen chunks. To make this possible, instead of representing a vector as a series pairs (`runlength,value`), we choose a representation based on prefix sums and store a series of (`position, value`) pairs (cf. Figure 6 (b) and (c)). In the position-based format, `values[i]` is the attribute's value of all tuples between `positions[i-1]` and `positions[i]-1` (for $i = 0$ the range is `0` to `positions[1]-1`). This layout consumes the same space (assuming that the same data type is used for run-lengths as for positions) and allows for scans almost as fast as the regular representation. The advantage, however, is that point accesses, that would require a scan in the other representation, can be sped up significantly through binary searches. In Section 5.6.2 we show a performance comparison.

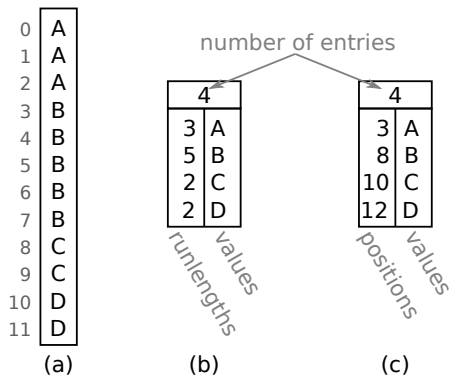

Figure 6: (a) Uncompressed chunk. (b) Regular RLE implementation (run-length RLE). (c) Position-based RLE.

Other common compression techniques are also conceivable, e.g. the reduction of the number of bytes used for a data type to what is actually required for the values in a chunk. In our implementation, however, we only integrate dictionary compression and run-length encoding into HyPer.

Abadi et al. [1] conclude that more heavy-weight compression schemes (such as Lempel-Ziff encoding) do not constitute a good trade-off between compression ratio and processing speed – even for a purely analytical database systems. Hence, compression schemes like dictionary compression and RLE seem to be a good choices for hybrid transactional and analytical systems as well.

## 3.5 Query Processing

There is a substantial amount of research about query processing and optimization of compressed data [32, 1, 14, 6].

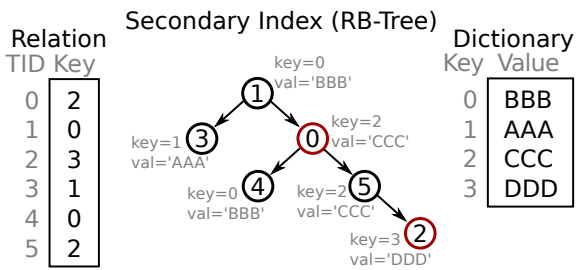

**Figure 7: Secondary tree index (gray information is not materialized). E.g. entry 1 compares less then entry 0, because the key at TID 1 ($0_{key}$) refers to a smaller value ($BBB$) as the key at TID 0 ($2_{key}$) which refers to $CCC$.**

Many of the insights from this work are also applicable to our approach. Order-preserving dictionary compression, however, can often be found in analytical database systems [2, 3] but is not feasible in write-intensive scenarios. In ordered dictionaries, $key_i < key_j$ implies that $value_i < value_j$. This property can be utilized to do query processing directly on the compressed representation of the data. While this property can speed up execution, it makes ordered dictionaries very difficult to maintain even in data warehouses, where data is inserted in bulk operations. Hybrid OLTP and OLAP systems have to handle high-frequency updates and inserts. Therefore, maintaining an ordered dictionary in this scenario is virtually impossible. However, we will present a novel variant of a secondary index. Using this index on the compressed attribute is an alternative that is feasible, outperforms an ordered dictionary in many cases and consumes significantly less memory than traditional indexes on string attributes.

Ordered dictionaries demonstrate their strength when executing queries with range and prefix filter conditions (e.g. `attribute LIKE 'prefix%'`). The currently proposed algorithm [3] first determines the qualifying keys through binary search in the dictionary. Then, the relation's attribute column is scanned and each key is tested for inclusion in the range. This algorithm is very efficient for unselective range queries. For more selective queries, however, a secondary tree index is dominant (as we show in Section 5.6.3) because it does not require a full table scan, but directly accesses the selected tuples. Figure 8 contrasts the two different processing techniques. The secondary index (tree index) can be constructed with very moderate memory consumption overhead: Instead of storing the values (e.g. strings) in the index and thus reverting the benefit of compression, we propose to only store the 8 byte TID of each tuple. The sort order of the TIDs is then determined by the order of the values they point to, as depicted in Figure 7. I.e. the index entry (TID) $tid_1$ compares less then entry $tid_2$, if the two keys $k_1$ and $k_2$ these TIDs point to refer to values in the dictionary $v_1$ and $v_2$ such that $v_1 < v_2$. If two values are equal, the TID serves as a tie-breaker in order to maintain a strict order. A strict order allows for efficient deletes and updates in the index as index entries of a given TID are quickly found. It is important to point out that, in contrast to pure OLAP systems, sorting or partitioning/cracking [17] the compressed vector is not possible in a hybrid OLTP and OLAP database system as it would require massive updates in the indexes that are required for the mission-critical transaction processing.

Navigating the tree index performs many random accesses. While this access pattern causes a performance problem for disk-based systems, in-memory systems can efficiently use this compact secondary index as an accelerator where prefix queries would benefit from order-preserving dictionary compression. We demonstrate the efficiency of this index in Section 5.6.3.

For equality filter conditions (e.g. `attribute = 'value'`), the scan operator first performs a lookup of the value in the dictionary (regardless of whether it is ordered or not) to determine the associated key and a subsequent scan of the compressed column passing only those tuples to the upstream operator that match the key. If a secondary index on the compressed attribute exists, the scan operator directly uses this index to determine the qualifying tuples and thus obviates a full table scan.

Filter conditions other than prefixes or equality comparisons cannot be efficiently processed with any of the techniques presented here. HyPer evaluates them by first scanning the dictionary and selecting the qualifying keys into a hash table. This hash table is then used for a hash join with the relation, i.e. it is probed with the keys of the compressed chunks. Since the hash table is likely to be small for many queries, it often fits into cache which makes this hash join very efficient.

All algorithms presented in this section are performed by the scan operator. The scan operator thus decompresses encoded chunks and unites tuples from compressed and from uncompressed chunks. Thereby, compression is transparent for all upstream operators and does not require their adaption.

## 4. IMPLEMENTATION DETAILS

### 4.1 Access Observer

The Access Observer component monitors reads and writes performed by OLTP threads in order to determine which parts of the database are cold and which ones are hot. We present different approaches for its implementation.

The first approach is purely software-based: Each OLTP thread records reads and writes itself. The advantage is that the granularity used to record accesses can be freely chosen. The obvious drawback is the significant overhead this imposes on the OLTP thread: Even when leaving out reads that are required to locate the requested values, a single invocation of TPC-C's NewOrder transaction performs over 50 read and over 100 write accesses on average. Since the purpose of the Access Observer and hot/cold clustering is to unburden the OLTP threads as much as possible, we dismiss this approach.

The second approach uses a technique often employed for live migration in virtual machine monitors like Linux' Kernel-based Virtual Machine (KVM) [20] and the Xen Virtual Machine Monitor [7]: The `mprotect` system call is used to prevent accesses to a range of virtual memory pages. When a read or write to this region occurs, a `SIGSEGV` signal is sent to the calling process, which has installed a signal handler that records the access and removes the protection from the page. This technique uses hardware-support, but still has a downside: For each page, the first write in an observation cycle causes a trap into the operating system. While this is no problem for transactions that mostly insert data, others that perform updates to distributed data items are impeded.

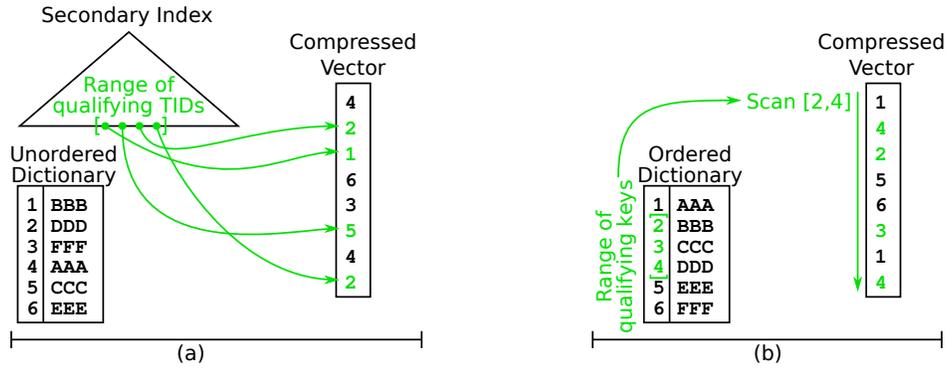

Figure 8: Execution of the range query ... BETWEEN 'BBB' AND 'DDD' using a (a) secondary index and (b) order-preserving dictionary.

The third approach also uses hardware-assistance, but has no overhead when monitoring the database. Here, the Access Observer component runs asynchronously to the OLTP (and OLAP) threads and uses information collected by the hardware.

Virtual memory management is a task modern operating systems master efficiently thanks to the support of the CPU's memory management unit (MMU). In particular, page frame reclamation relies on hardware assistance: The MMU sets flags for each physical memory page indicating if addresses within this page have been accessed (`young`) or modified (`dirty`). The Linux Virtual Memory Manager uses this information during page frame reclamation to assess if a page is in frequent use and whether it needs to be written to the swap area before a different virtual page can be mapped on it [13]. In HyPer, we prevent memory pages from getting paged out to the swap area by using the `mlock` system call. Thus we can read and reset the `young` and `dirty` flags in each observation cycle to monitor accesses to the database with virtually no overhead.

Figure 9 shows the interaction of different components of HyPer. We have implemented this type of Access Observer as a kernel module for an (unmodified) Linux kernel for the x86 architecture. On other architectures, Linux even provides a dirty bit that can be used exclusively by user space programs [12]. While the component is certainly platform-dependent, Linux is not the only system where this approach is conceivable. To broaden the platform support for database systems using the Access Observer, the `mprotect`-based approach can be used as a fall-back option.

Since the kernel module can distinguish between read-only and write accesses, we can refine our hot/cold model from Section 3.3. Compression techniques, like dictionary compression, that incur little overhead on read-only point accesses can be applied to cold chunks that are still read, but not written. Compression schemes where point-accesses are more expensive can be applied once the chunk is neither read nor written to.

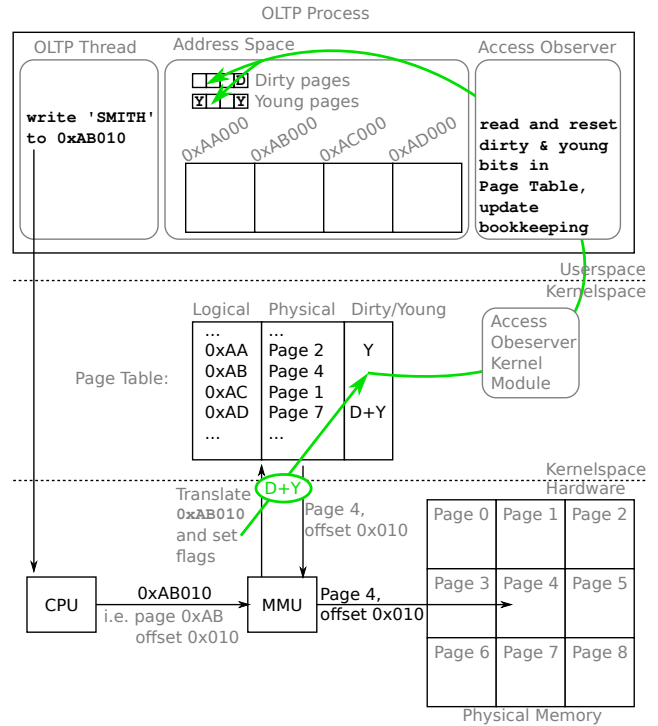

Figure 9: Access Observer architecture of the third approach (simplified illustration without TLB, caches, etc.).

The hardware-supported approaches operate on page granularity. Thus by themselves they cannot distinguish between a single change and many changes in one page. For those (small) parts of the data, where this distinction is important, the hardware-supported techniques could be combined with the software-based approach. In our test-cases, however, we did observe a situation where this was necessary.

For hot/cold clustering, solely accesses from OLTP threads should be take into account. While OLAP queries never perform writes, they frequently scan over entire relations. In both hardware-based approaches, this causes the page to be considered `young`. Thus, read flags of a given relation are only considered, if no OLAP query has accessed this relation in the last observation cycle.

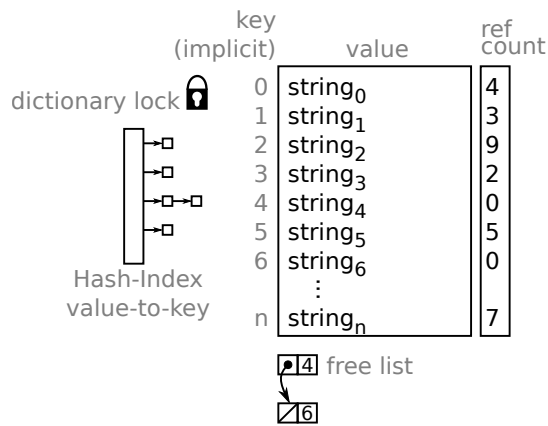

Figure 10: Dictionary data structure for strings.

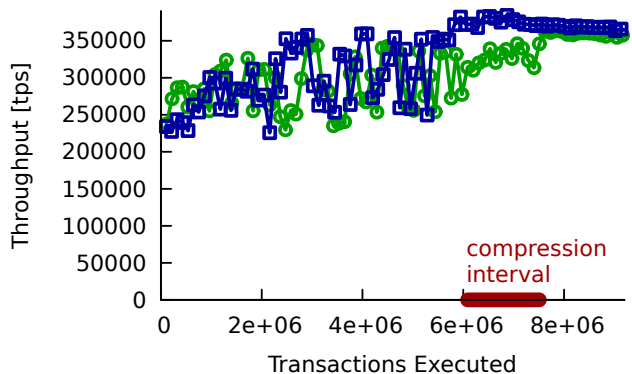

Figure 11: Transactional performance with activated compression (○) and without compression (□).

### 4.2 Dictionary Compression

Dictionaries consist of two columns: A reference counter indicating the number of tuples that reference the value and the actual value. The key is not stored explicitly, but is the offset into the dictionary, as shown in Figure 10. A reference count of zero indicates that this slot in the dictionary can be reused. In order to avoid duplicate entries, a value-to-key hash-index is used to lookup existing values when a new tuple is inserted or a compressed value is being updated.

We maintain one dictionary per relation. Following the transaction model described in Section 3.2, this means that multiple OLTP threads access the dictionary concurrently. Since only cold data is dictionary-compressed, we expect very little updates in the dictionary by OLTP threads. Yet, an OLTP thread may make modifications in a vector that is currently being compressed. Therefore, during the compression of a chunk, the reorganization thread uses the Access Observer to track write accesses, while a new (and smaller) vector is filled with the dictionary keys equivalent to the original values. If a memory page was modified during compression and processing a value within this page has already caused changes in the dictionary, it is re-worked: The modified page is scanned and every value is compared to the value the new key-vector points to: If the values do not match, the dictionary's reference counter for the value pointed to by the current key is decremented (i.e. the insert into the dictionary is undone) and the new value is being compressed. This optimistic concurrency control is facilitated by the Access Observer's ability to detect writes retrospectively.

As OLTP and reorganization threads access the dictionary concurrently, they need to synchronize dictionary access with a lock. However, since the hot part of the data is uncompressed, transactions inserting new tuples or updating/deleting hot tuples never have to acquire the dictionary lock. Therefore lock contention is not a problem here.

## 5. EVALUATION

In this section, we substantiate our claims that transactional data can be compressed with very little overhead and that the performed compression is beneficial for query processing. Basis for the experiments is the CH-BenCHmark [8], a combination of the TPC-C with TPC-H-like queries operating on the same data concurrently. The schema is an entirely unmodified TPC-C schema extended by three fixed-size relations from TPC-H: SUPPLIER, NATION and REGION. The transactional part of the workload consists of the five original TPC-C transactions (New-Order, Payment, Order-Status, Delivery, Stock-Level), with the following changes:

- Since the TPC-C generates strings with high entropy that exhibits little potential for compression and is unrealistic in business applications, we replaced the strings with US Census 2010 data. E.g. for the attribute OL_DIST_INFO we do not use a random string, but a last name (e.g. the last name of the person responsible for this orderline). The last name is selected from the list of the most common 88.800 family names with a probability according to the frequency of the name.

- The CH-BenCHmark follows the proposal made by VoltDB [31] to deviate from the underlying TPC-C benchmark by not simulating the terminals and by generating client requests without any think-time.

We conduct our experiments on a server with two quad-core Intel Xeon CPUs clocked at 2.93GHz and with 64GB of main memory running Redhat Enterprise Linux 5.4. The benchmark is scaled to 12 warehouses.

### 5.1 Transactional Performance

We quantify the impact of our compression technique on transaction throughput using the transactional part of the benchmark. In the following experiment, we compare benchmark runs with two different HyPer setups: One without any compression techniques and one including the compression techniques. In both runs, we configured HyPer to use five OLTP threads.

The CH-BenCHmark schema has 32 attributes of type `char` or `varchar` that have length 20 or more and thus exhibit compression-potential. The most interesting relations, however, are the three constantly growing relations out of which two, ORDERLINE and HISTORY, have a `varchar` attribute of length 24 and thus require recurring compression of freshly inserted tuples.

Figure 11 shows the result of the CH-BenCHmark runs in which we simulate one week of transaction processing of the world's larges online retailer: Amazon generates a yearly revenue of around $30 billion, i.e. assuming an average item price of $30, Amazon adds nearly 20 million orderlines to its database each week. We configured HyPer to compress of the ORDERLINE and HISTORY relations' cold chunks containing the same amount of transactional data Amazon generates in one week (according to our back-of-the-envelope calculation) all at once, to show the impact of the compression as clearly as possible. In this compression interval, HyPer compressed 18.3 million ORDERLINE and 0.9 million HISTORY tuples in 3.8 seconds. The transaction throughput in these 3.8s was 12.9% slower in the setup that performed compression than in the setup without compression. Since there are no synchronization points for the compression thread and the OLTP threads while a chunk is being compressed, this seems to result from competing accesses to the memory bus. Note that fluctuations before the compression interval do not result from the techniques described in this paper, but originate from the benchmark driver.

## 5.2 Update of Cooling and Frozen Data

The Access Observer tries to ensure that the number of accesses to cool and frozen chunks is minimal and our experiments with the CH-BenCHmark/TPC-C show that this goal is often achievable. However, individual accesses to these tuples cannot be ruled out in general, so we quantify their costs here.

Cooling chunks are diagnosed by the Access Observer to only contain a few tuples that change. Thus HyPer chooses to relocate these tuples to a hot chunk in order to be able to compress the cooling chunk. We mark hot chunks "cooling" in order to quantify the costs of these relocations. In a relation with 50 million tuples, we forcefully mark all chunks "cooling" and then update them, to trigger their relocation to a newly created hot chunk. With 3,595ms, the run time is over twice as long as the updates take when all chunks are correctly marked "hot" (1,605ms) and thus require no relocation of tuples. This result indicates that while it requires extra processing to access a cooling tuple, the amortized cost over all accesses is negligible, given the fact that accesses to cooling chunks are rare compared to accesses to hot chunks. If massive accesses to cooling chunks should occur, the Access Observer detects the warming up of the chunk and switches its temperature back to hot, which prevents further relocations.

Frozen chunks are compressed and reside on huge memory pages and their tuples are therefore not modified in place, but invalidated. In addition to relocating the tuple as for cooling chunks, it also requires an update in the partition's invalidation status. Unlike cooling chunks, frozen chunks *must* perform this procedure. We first perform a run where we update all 50 million tuples in sequence which requires about the same time (3,436ms) as the updates in the cooling chunks. I.e. the costs of invalidating the tuples are dominated by the costs of relocating them, as only one invalidation entry per chunk is being created. When updating random tuples, inserts into the validation status are more costly: We update 10,000 random orders in the ORDERLINE relation. Each order consists of 10 consecutively located orderline tuples, i.e. 100,000 orderlines are updated. Performing these updates in frozen chunks takes 46ms, while it takes 18ms in hot chunks. As updates in frozen chunks are rare, this slowdown of factor 2.55 is unproblematic.

| Scale | Instant Compr. | No Compr. |
|---|---:|---:|
| 10M orderlines $2^{15}$ unique values | 4,249ms | 2,790ms |
| 10M orderlines $2^{18}$ unique values | 5,589ms | 2,791ms |
| 50M orderlines $2^{15}$ unique values | 19,664ms | 12,555ms |
| 50M orderlines $2^{18}$ unique values | 26,254ms | 12,614ms |

Table 1: The costs of instant compression: Time required to insert 10M and 50M orderlines.

## 5.3 Access Observer

In this section, we compare the impact of the Access Observer implementation on HyPer's performance. While the implementation based on the `young` and `dirty` flags imposes absolutely no overhead on the OLTP threads, we measured the impact of the alternative implementation (using `mprotect`) by protecting vectors of `numeric` attributes and then performing random writes to them. This situation arises multiple times in TPC-C, e.g. in the stock or customer relation. When performing 100,000 updates to 10 million entries, the `mprotect`ed version requires 94ms, while a run without takes only 4ms. Consequently, the no-overhead implementation based on the `young` and `dirty` flags is our first choice for HyPer.

## 5.4 Compression Performance

We conduct experiments to substantiate the claim that our approach of lazy-compression is superior to eager compress. For run-length encoding, the disadvantages of instantly compressing tuples are obvious: Updates of RLE values can split a run and generate up to two additional entries that do not fit in the vector (and cause the relocation of all following entries if the do). In addition, locating an attribute's value associated with a TID is possible due to our design, but still compromises performance. Therefore we limit the experiment to dictionary compression. Table 1 shows the time required to insert TPC-C orderlines when compressing instantly and when performing no compression of the `char(24)` attribute OL_DIST_INFO. Thus, the additional lookup/insert in the dictionary can slow down the insert rate to 50%. This results from the fact that without compression, inserting an orderline only requires an insert into the primary key index in addition to actually inserting the 10 attributes and is therefore extremely fast.

## 5.5 Compression Effectiveness

While the focus of this work is on the integration of compression techniques rather than on compression itself and thus has only limited influence on the compression factor, we measure how much our approach can reduce the memory consumption in the CH-BenCHmark. We execute transactions until our database size is 50GB and then activate our compression feature to compress the cold parts of the database. We configure HyPer to use dictionary compression for all cold vectors of string attributes. We populated these attributes with values according to a Zipf distribution with parameter $s = 1.2$. This yields a compression ratio of 1.12. This low number does not result from a possible ineffectiveness in our design, but from the fact that only two attributes in the three continuously growing relations have character types (`h_data` and `ol_dist_info`) and their length is only 24 characters. When increasing the length to 240 characters, the compression factor rises to 2.42.

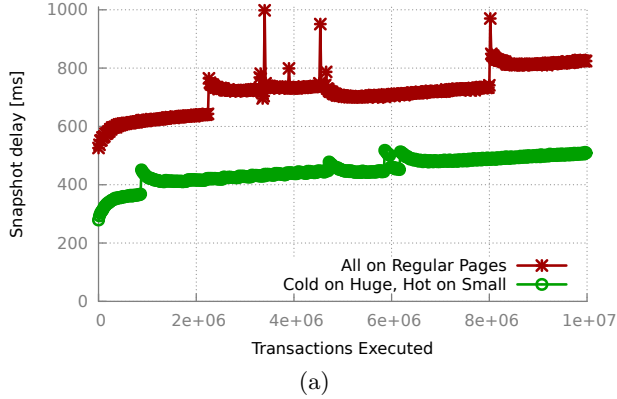
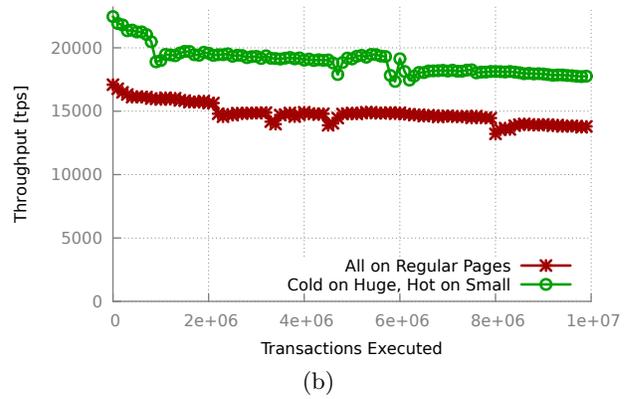

Figure 12: Impact of page size on snapshot performance (a) and transactional performance (b).

Activating run-length encoding in addition to dictionary compression shrinks ORDERLINE's `ol_o_id` attribute by a factor of 3.3 as each order has an average of 10 orderlines that are stored consecutively and thus produce runs of average length 10. The timestamp attributes `o_entry_d` and `h_date` are naturally sorted in the database, as they are set to the current date and time on insert. Thus they contain runs of several thousand entries, resulting in extreme compression factors for these attributes.

For all three growing relations, HISTORY, ORDERLINE and ORDER, the Access Observer indicates that all chunks but the one or two last ones could be frozen at any time during the benchmark run.

## 5.6 Query Performance

Query performance in hybrid OLTP and OLAP systems depends on two factors: Snapshot performance and query execution time. Compression and the usage of huge pages for frozen chunks improve snapshot performance. Query execution time benefits from huge pages and compression, too. Scanning invalidated tuples can have a negative effect on query performance, but we demonstrate that this drawback is negligible.

### 5.6.1 Snapshot Performance

Snapshot performance primarily depends on the size of the page table that has to be copied when a snapshot is created. It impacts transaction throughput as well as average query response times: During the creation of a snapshot, OLTP workers must be quiesced and queries waiting for the snapshot are stalled. We compare a setup were the whole database resides on regular-sized pages to a configuration in which huge pages are used for frozen chunks.

Figure 12 compares both setups when a new snapshot is created every 40,000 transactions while a single OLTP thread processes transactions. Plot (a) shows the benefits of using huge pages for frozen chunks for snapshot creation performance. Plot (b) demonstrates that the use of huge pages improves transaction throughput.

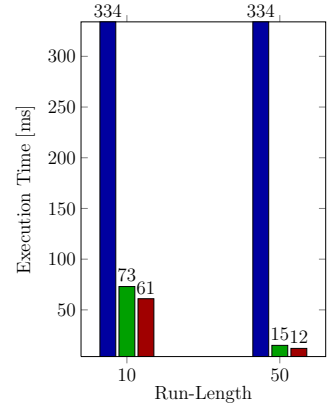

Figure 13: Comparison of RLE schemes when summing up 500M values. ■ uncompressed, ■ position-based RLE (prefix sum), ■ regular RLE.

### 5.6.2 Run-Length Encoding

In Section 3.4, we described the layout of a RLE vector that uses positions instead of run-lengths. We did so to bridge the gap between scan-based OLAP accesses and occasional point-wise OLTP accesses. Figure 13 shows the execution time of queries summing up 500 million integers for two different run-lengths. The position-based encoding is hardly slower than the regular run-length encoding, when comparing both runs on compressed data with the run on uncompressed data.

The benefits of this trade-off in scan performance are relatively efficient point-accesses: For the test-case with run-length 10, lookups using our position-based RLE were only 7 times slower than lookups in uncompressed data, for the test case with run-length 50, the factor is 5 times. Point accesses in regular run-length encoded data on the other hand require a linear scan of all runs until the TID is found, resulting in up to six orders of magnitude longer lookup times in our benchmark. Thus the use of regular RLE is out of the question, even though point accesses to compressed data are infrequent. Position-based RLE however seems to be an excellent compromise.

### 5.6.3 Prefix Scan Performance

CH-BenCHmark's query $Q_1$ performs a scan of the ORDERLINE relation, groups and sorts by the orderline number and computes different aggregations. Since our approach

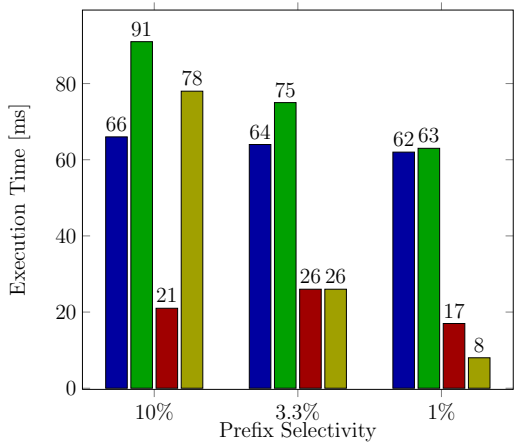

**Figure 14: Comparison of range scan algorithms executing $Q_1$ on 3.6M orderlines.** ■ uncompressed, ■ unordered dictionary (i.e. hash-join with dictionary), ■ ordered dictionary, ■ secondary index.

```
SELECT  ol_number,
        SUM(ol_quantity) AS sum_qty,
        SUM(ol_amount) AS sum_amount,
        AVG(ol_quantity) AS avg_qty,
        AVG(ol_amount) AS avg_amount,
        COUNT(*) AS count_order
FROM    orderline
WHERE   ol_delivery_d>'2007-01-02 00:00:00'
        AND ol_dist_info LIKE '<prefix>%'
GROUP BY ol_number ORDER BY ol_number
```

**Figure 15: CH-BenCHmark Query $Q_1$ with additional (prefix) filter condition ol_dist_info LIKE '<prefix>%'.**

only changes the scan operator, we conduct experiments with variations of this scan-oriented query to carve out the performance impact of dictionary compression.

Figure 14 shows the performance of different prefix scan algorithms on the query listed in Figure 15. The attribute OL_DIST_INFO of the ORDERLINE relation is populated with the aforementioned US family name dataset to obtain a realistic distribution of values.

Since HyPer pushes simple selection conditions into the scan operator, the following algorithms are possible: The baseline for the experiment is an algorithm that is not only applicable to prefix queries, but to arbitrary filter conditions: It scans the dictionary and selects all matches into a hash table which is then used to probe the attribute's dictionary keys into. The algorithm's performance mostly depends on the size of the relation, but also on whether or not the hash table fits into cache.

To evaluate the performance of order-preserving dictionaries, we use the algorithm described in Section 3.4 that is based on the determination of the qualifying key-range and a subsequent scan of the relation. This algorithm's performance is again mostly determined by the size of the relation, but the probing phase is far less costly compared to the first algorithm which requires the computation of a hash value and a lookup in the hash table. The selection of the key-range in the dictionary is also faster then the creation of the hash table in the first algorithm. However, the time spent in the dictionary is dominated by the probe phase in both algorithms, as it accounts for less then 1% of the run-time.

We compare these two algorithms with a lookup in a secondary index. The red-black-tree index we used in our experiment does not store the values of the indexed attribute, but only the TIDs referencing the dictionary-keys of these values (cf. Figure 8). Thus it is very memory consumption friendly. Our red-black-tree requires 24 bytes per entry, but a B-tree could be used instead and would consume only a little bit more than the 8 byte payload per entry (mostly depending on the B-tree's load-factor).

Figure 14 shows that for selectivities of 3.3% or higher, the ordered dictionary is faster than the direct lookup in the secondary index. For selectivities of less than 3.3%, the secondary index outperforms the order-preserving dictionary. As analytical queries often have single-digit selectivities, we believe that secondary indexes are a valid alternative for ordered-dictionaries, especially in high-frequency update and insert scenarios where ordered-dictionaries are impossible to maintain efficiently.

To demonstrate that our invalidation mechanism for frozen chunks hardly affects scan performance, we invalidate 10,000 random orders, i.e. about 100,000 orderlines in the dataset (as every order has an average of 10 orderlines) and compare the scan results with a run in which the same amount of orderlines where simply deleted. This means that in both setups, 3.5M orderlines are aggregated by query $Q_1$. Even in this extreme scenario where as much as $\frac{1}{36}$ of the dataset is invalid and the invalidations are scattered, the performance decline of the three prefix queries from Figure 14 was only between 4.5% and 7.9%.

## 6. CONCLUSION

We have shown that compression can – against common belief – indeed be integrated into high-performance OLTP systems without impacting transaction throughput. This can be achieved by relieving the OLTP threads from all compression-related tasks and performing compression asynchronously as soon as tuples are stopped being used by transactions. We have presented hardware-assisted, low-overhead monitoring techniques to assess when tuples should be compressed. This, again, unburdens transaction processing from the overhead of keeping track of all data accesses. Experiments with TPC-C have shown that our hot/cold clustering technique has very little overhead and can identify the bulk of the database as cold. Cold parts of the database indentified through the aforementioned observation method can be easily compressed using various compression schemes and can be physically optimized for query processing and compact snapshots by freezing them and relocating them to huge memory pages. Future work includes other optimizations such as small materialized aggregates for each frozen chunk. We have also shown the impact of compression on query performance and have presented a memory-consumption friendly secondary index for the efficient evaluation of unselective prefix and range queries.

## Acknowledgements

This work is partially funded by the German Research Foundation (DFG).

We would like to thank IBM for supporting Florian Funke with an IBM Ph.D. Fellowship Award.

# 7. REFERENCES

[1] D. J. Abadi, S. Madden, and M. Ferreira. Integrating Compression and Execution in Column-Oriented Database Systems. In *SIGMOD*, pages 671–682, 2006.

[2] G. Antoshenkov, D. B. Lomet, and J. Murray. Order Preserving Compression. In *ICDE*, pages 655–663, 1996.

[3] C. Binnig, S. Hildenbrand, and F. Färber. Dictionary-based Order-preserving String Compression for Main Memory Column Stores. In *SIGMOD*, pages 283–296, 2009.

[4] P. A. Boncz, M. Zukowski, and N. Nes. MonetDB/X100: Hyper-Pipelining Query Execution. In *CIDR*, pages 225–237, 2005.

[5] T. Cao, M. A. V. Salles, B. Sowell, Y. Yue, A. J. Demers, J. Gehrke, and W. M. White. Fast Checkpoint Recovery Algorithms for Frequently Consistent Applications. In *SIGMOD*, pages 265–276, 2011.

[6] Z. Chen, J. Gehrke, and F. Korn. Query Optimization In Compressed Database Systems. In *SIGMOD*, pages 271–282, 2001.

[7] C. Clark, K. Fraser, S. Hand, J. G. Hansen, E. Jul, C. Limpach, I. Pratt, and A. Warfield. Live Migration of Virtual Machines. In *NSDI*, pages 273–286, 2005.

[8] R. Cole, F. Funke, L. Giakoumakis, W. Guy, A. Kemper, S. Krompass, H. A. Kuno, R. O. Nambiar, T. Neumann, M. Poess, K.-U. Sattler, M. Seibold, E. Simon, and F. Waas. The mixed workload CH-benCHmark. In *DBTest*, page 8, 2011.

[9] G. P. Copeland and S. Khoshafian. A Decomposition Storage Model. In *SIGMOD*, pages 268–279, 1985.

[10] P. J. Denning. The Working Set Model for Program Behaviour. *Communications of the ACM*, 11(5):323–333, 1968.

[11] F. Färber, S. K. Cha, J. Primsch, C. Bornhövd, S. Sigg, and W. Lehner. SAP HANA Database: Data Management for Modern Business Applications. *SIGMOD Record*, 40(4):45–51, 2011.

[12] F. Funke. [s390] introduce dirty bit for kvm live migration. Patch integrated in Linux kernel 2.6.28, Oct 2008. http://git.kernel.org/?p=linux/kernel/git/stable/linux-stable.git;a=commit;h=15e86b0c752d50e910b2cca6e83ce74c4440d06c.

[13] M. Gorman. *Understanding the Linux Virtual Memory Manager*. Prentice Hall PTR, 2004.

[14] G. Graefe and L. D. Shapiro. Data Compression and Database Performance. In *In ACM/IEEE-CS Symposium On Applied Computing*, pages 22–27, 1991.

[15] S. Héman, M. Zukowski, N. J. Nes, L. Sidirourgos, and P. A. Boncz. Positional Update Handling in Column Stores. In *SIGMOD*, pages 543–554, 2010.

[16] A. L. Holloway, V. Raman, G. Swart, and D. J. DeWitt. How to Barter Bits for Chronons: Compression and Bandwidth Trade Offs for Database Scans. In *SIGMOD*, pages 389–400, 2007.

[17] S. Idreos, M. L. Kersten, and S. Manegold. Database Cracking. In *CIDR*, pages 68–78, 2007.

[18] R. Kallman, H. Kimura, J. Natkins, A. Pavlo, A. Rasin, S. B. Zdonik, E. P. C. Jones, S. Madden, M. Stonebraker, Y. Zhang, J. Hugg, and D. J. Abadi. H-Store: A High-Performance, Distributed Main Memory Transaction Processing System. *PVLDB*, 1(2):1496–1499, 2008.

[19] A. Kemper and T. Neumann. HyPer: A Hybrid OLTP&OLAP Main Memory Database System Based on Virtual Memory Snapshots. In *ICDE*, pages 195–206, 2011.

[20] A. Kivity, Y. Kamay, D. Laor, U. Lublin, and A. Liguori. kvm : the Linux Virtual Machine Monitor. kernel.org, 2007.

[21] J. Krüger, M. Grund, C. Tinnefeld, H. Plattner, A. Zeier, and F. Faerber. Optimizing Write Performance for Read Optimized Databases. In *Database Systems for Advanced Applications*, pages 291–305. Springer, 2010.

[22] J. Krüger, C. Kim, M. Grund, N. Satish, D. Schwalb, J. Chhugani, H. Plattner, P. Dubey, and A. Zeier. Fast Updates on Read-Optimized Databases Using Multi-Core CPUs. *PVLDB*, 5(1):61–72, 2012.

[23] R. A. Lorie. Physical Integrity in a Large Segmented Database. *TODS*, 2(1):91–104, 1977.

[24] S. Manegold, P. A. Boncz, and M. L. Kersten. What Happens During a Join? Dissecting CPU and Memory Optimization Effects. In *VLDB*, pages 339–350, 2000.

[25] H. Mühe, A. Kemper, and T. Neumann. How to Efficiently Snapshot Transactional Data: Hardware or Software Controlled? In *DaMoN*, pages 17–26, 2011.

[26] T. Neumann. Efficiently Compiling Efficient Query Plans. In *VLDB*, pages 539–550, 2011.

[27] Oracle. Advanced Compression with Oracle Database 11g. *Whitepaper*, 2011.

[28] V. Raman, G. Swart, L. Qiao, F. Reiss, V. Dialani, D. Kossmann, I. Narang, and R. Sidle. Constant-Time Query Processing. In *ICDE*, pages 60–69, 2008.

[29] Transaction Processing Performance Council. TPC-C specification. www.tpc.org/tpcc/spec/TPC-C\_v5-11.pdf, 2010.

[30] VoltDB. Technical Overview. http://www.voltdb.com, March 2010.

[31] VoltDB Community. VoltDB TPC-C-like Benchmark Comparison-Benchmark Description. http://community.voltdb.com/node/134, 2010.

[32] T. Westmann, D. Kossmann, S. Helmer, and G. Moerkotte. The Implementation and Performance of Compressed Databases. *SIGMOD Record*, 29(3):55–67, 2000.

[33] M. Zukowski, S. Héman, N. Nes, and P. A. Boncz. Super-Scalar RAM-CPU Cache Compression. In *ICDE*, page 59, 2006.